\documentclass[conference,10pt]{IEEEtran}
\IEEEoverridecommandlockouts

\usepackage[letterpaper,top=0.75in,bottom=1.09in,left=0.635in,right=0.635in]{geometry}
\usepackage{amsmath,amssymb,amsfonts}
\usepackage{graphicx}
\usepackage{xcolor}
\usepackage{cite}
\usepackage{url}
\usepackage{algorithm}
\usepackage{algpseudocode}
\usepackage[nolist]{acronym}
\usepackage{tikz}
\usetikzlibrary{arrows.meta,positioning}

\begin{document}

\title{Coherent Multiband OFDM Sensing via Low-Complexity Gap Reconstruction
\thanks{This work was supported by the CNIT National Laboratory of Wireless Communications (WiLab) and the WiLab-Huawei Joint Innovation Center.}
}
\author{\IEEEauthorblockN{
Lorenzo~Pucci\IEEEauthorrefmark{1},
Leonardo~Pucci\IEEEauthorrefmark{2},
and Andrea~Giorgetti\IEEEauthorrefmark{1}\IEEEauthorrefmark{2}
}
%\IEEEauthorblockA{Wireless Communications Laboratory, CNIT, DEI, University of Bologna, Italy\\ 
%Email: \{tommaso.bacchielli2, lorenzo.pucci3, andrea.giorgetti\}@unibo.it\\
\IEEEauthorblockA{
    \IEEEauthorrefmark{1}National Laboratory of Wireless Communications (WiLab), CNIT, Italy\\
    \IEEEauthorrefmark{2}DEI, University of Bologna, Italy\\
    Emails: lorenzo.pucci@wilab.cnit.it, leonardo.pucci@studio.unibo.it, andrea.giorgetti@unibo.it
    }
}

\maketitle

\begin{acronym}
%%%% A
\acro{ADC}{analog-to-digital converter}
\acro{AF}{ambiguity function}
\acro{AI}{artificial intelligence}
\acro{ALS}{alternating least square}
\acro{AWGN}{additive white Gaussian noise}
\acro{AoA}{angle of arrival}
\acro{AoD}{angle of departure}
\acro{AP}{access point}
%%%% B
\acro{BER}{bit error rate}
\acro{BF}{beamformer}
\acro{BS}{base station}
\acro{BP}{back-projection}
%%%% C
\acro{CPE}{common phase error}
\acro{CA}{carrier aggregation}
\acro{CAF}{cross-ambiguity function}
\acro{CC}{carrier component}
\acro{CCC}{cyclic cross-correlation}
\acro{CFR}{channel frequency response}
\acro{CFO}{carrier frequency offset}
\acro{CP}{cyclic prefix}
\acro{CIR}{channel impulse response}
\acro{CPI}{coherent processing interval}
\acro{CRLB}{Cram\'{e}r-Rao lower bound}
\acro{CRB}{Cram\'{e}r-Rao bound}
\acro{CSI}{channel state information}
\acro{CS}{compressive sensing}
\acro{ch.f.}{characteristic function}
%%%% D
\acro{DMRS}{demodulation reference signal}
\acro{DoA}{direction of arrival}
\acro{DoD}{direction of departure}
\acro{DFRC}{dual-function(al) radar communications}
\acro{DFT}{discrete Fourier transform}
\acro{DPSK}{differential phase shift keying}
\acro{DSSS}{direct sequence spread spectrum}
\acro{DL}{downlink}
%%%% E
\acro{EIRP}{effective isotropic radiated power}
\acro{EKF}{extended Kalman filter}
\acro{ELP}{equivalent low-pass}
\acro{EMSE}{expectation of the MSE}
%%%% F
\acro{FAR}{false alarm rate}
\acro{FC}{fusion center}
\acro{FR}{frequency range}
\acro{FR3}{frequency range 3}
\acro{FR2}{frequency range 2}
\acro{FFT}{fast Fourier transform}
\acro{FDD}{frequency-division duplexing}
\acro{FIM}{Fisher information matrix}
\acro{FMCW}{frequency modulated continuous wave}
\acro{FRT}{fraction of resolved targets}
%%%% G
\acro{GOSPA}{generalized optimal sub-pattern assignment}
\acro{GDOP}{geometric dilution of position}
\acro{GLRT}{generalized likelihood ratio test}
\acro{GPU}{graphics processing unit}
\acro{GSM}{global system for mobile communications}
%%%% H
%%%% I
\acro{ISAC}{integrated sensing and communication}
\acro{IFFT}{inverse fast Fourier transform}
\acro{IDFT}{inverse discrete Fourier transform}
\acro{IMM}{interacting multiple model}
\acro{IVA}{inverse virtual aperture}
\acro{i.i.d.}{independent, identically distributed}
\acro{ISAFS}{iterative SAF subtraction}
\acro{IPP}{image projection plane}
\acro{IRCI}{inter-range-cell interference}
\acro{ISI}{inter-symbol interference}
\acro{ISAC}{integrated sensing and communication}
\acro{ICI}{intercarrier interference}
\acro{IC}{image contrast}
%%%% J
\acro{JSC}{joint sensing and communication}
\acro{JCAS}{joint communication radio/radar sensing}
\acro{JRC}{joint radar (and) communications}
\acro{JCR}{joint communications (and) radar}
\acro{JCAVA}{joint communication and VA imaging}
%%%% K
%%%% L
\acro{LoS}{line-of-sight}
\acro{LO}{local oscillator}
\acro{LRT}{likelihood ratio test}
\acro{LS}{least square}
\acro{LTE}{long term evolution}
\acro{LFM}{linear frequency modulation}
%%%% M
\acro{MB}{multiband}
\acro{MC}{Monte Carlo}
\acro{MIMO}{multiple-input multiple-output}
\acro{MUSIC}{MUltiple SIgnal Classification}
\acro{MDL}{minimum description length}
\acro{MI}{mutual information}
\acro{MF}{matched filter}
\acro{ML}{maximum likelihood}
\acro{MLE}{maximum likelihood estimator}
\acro{MSE}{mean-square error}
\acro{MAC}{medium access control}
\acro{MU}{multi-user}
\acro{MUSIC}{multiple signal classification}
%%%% N
\acro{NR}{new radio}
\acro{NN}{neural network}
\acro{NRPP}{new radio positioning protocol}
\acro{NLoS}{non-line-of-sight}
%%%% O
\acro{OFDM}{orthogonal frequency division multiplexing}
\acro{OFDMA}{orthogonal frequency division multiple access}
\acro{OSPA}{optimal subpattern assignment}
\acro{OTA}{over-the-air}
\acro{OMP}{orthogonal matching pursuit}
%%%% P
\acro{PAPR}{peak-to-average power ratio}
\acro{PDF}{probability density function}
\acro{PFA}{polar formatting algorithm}
\acro{PSD}{power spectral density}
\acro{PEB}{position error bound}
\acro{PMCW}{phase modulated continuous wave}
\acro{PO}{phase offset}
\acro{PGA}{phase gradient autofocus}
\acro{PCA}{principal component analysis}
%%%% Q
\acro{QPSK}{quadrature phase shift keying}
\acro{PSK}{phase-shift keying}
\acro{p.d.f.}{probability density function}
%%%% R
\acro{RCS}{radar cross-section}
\acro{RF}{radio frequency}
\acro{RMSE}{root mean square error}
\acro{RMC}{rotation motion compensation}
\acro{RoI}{region of interest}
\acro{RPO}{random phase offset}
\acro{r.v.}{random variable}
\acro{RCMC}{range cell migration correction}
\acro{RDA}{range-Doppler algorithm}
\acro{Rx}{receiver}
\acro{RMS}{root mean square}
%%%% S
\acro{SAR}{synthetic aperture radar}
\acro{SSB}{single-subband}
\acro{SRI}{sensing repetition interval}
\acro{SCM}{sample covariance matrix}
\acro{SDR}{software defined radio}
\acro{SI}{self interference}
\acro{SISO}{single input single output}
\acro{SNR}{signal-to-noise ratio}
\acro{SU}{single-user}
\acro{SVD}{singular value decomposition}
\acro{SPEB}{squared position error bound}
\acro{SMI}{standard multistatic imaging}
\acro{SWMP}{swath-width matched pulse}
%%%% T
\acro{ToA}{time of arrival}
\acro{TO}{timing offset}
\acro{TDD}{time-division duplexing}
\acro{TDOA}{time difference of arrival}
\acro{TLRS}{target localisation reference signal}
\acro{TPU}{tensor processing unit}
\acro{TMC}{translational motion compensation}
\acro{Tx}{transmitter}
\acro{TRP}{transmit receive point}
%%%% U
\acro{UAV}{unmanned aerial vehicle}
\acro{UE}{user equipment}
\acro{ULA}{uniform linear array}
\acro{UWB}{ultra-wideband}
\acro{USRP}{universal software radio peripheral}
\acro{U6G}{upper 6 GHz}
%%%% V
\acro{V2V}{vehicle-to-vehicle}
\acro{VA}{virtual aperture}
%%%% W
\acro{WLS}{weighted least square}
%%%% X
%%%% Y
%%%% Z
\acro{ZZLB}{Ziv-Zakai lower bound}
\acro{ZZB}{Ziv-Zakai bound}
\end{acronym}
\begin{abstract}
This paper investigates coherent multiband \ac{OFDM} sensing within an \ac{ISAC} framework. We consider an intra-band configuration in which two sensing subbands of equal width are allocated symmetrically within the same \ac{OFDM} channel, while the central portion remains available for communication. We address the reconstruction of missing frequency-domain samples induced by the spectral gap and the suppression of the resulting grating lobes in the delay profile. To this end, we propose a low-complexity iterative reconstruction method consisting of an initial delay-domain equalization stage and an iterative apodization-based operator with data-consistency enforcement. Performance results for multi-target scenarios show that the proposed approach remains close to the full-band reference for moderate gap sizes and degrades only for larger gaps because of residual grating lobes. Compared with the compressed-sensing-based \ac{OMP} baseline, it exhibits a more favorable performance trend as the number of targets increases, especially in the practically relevant low-\ac{SNR} regime, while offering a complexity scaling that is independent of the estimated number of targets.
\end{abstract}

%\begin{IEEEkeywords}
%OFDM sensing, intra-band sensing, carrier aggregation, complex apodization, signal reconstruction, ISAC.
%\end{IEEEkeywords}

\acresetall
% ============================================================
\section{Introduction}
% 1) Context: ISAC + OFDM, sensing uses subcarriers
% 2) Motivation: intra-channel multiband sensing (two disjoint sensing subbands) with a comm-only gap
% 3) Challenge: zero-filled missing band => range ghosts / grating lobes, harming detection & resolution
% 4) Existing approaches: compressed sensing / OMP + model-order (MDL), computationally heavy; requires assumptions
% 5) Our approach: optics-inspired complex apodization for grating lobes (not classical sidelobes), iterative reconstruction
% 6) Contributions (bullet list)
% \begin{itemize}
%   \item \textbf{Problem formulation:} OFDM sensing with two disjoint subbands and a null sensing gap within a single channel; show grating-lobe mechanism.
%   \item \textbf{Method:} (i) range-domain equalization kick-start using cross-band consistency, (ii) iterative complex apodization-like operator + data consistency to rebuild missing spectrum.
%   \item \textbf{No model-order / low complexity:} FFT-based iterations; no need to estimate number of targets.
%   \item \textbf{Evaluation:} single-target ghost suppression and convergence; gap sweep (up to 0.5) with \ac{IGLR}/PSLR-like metrics; two-target resolution analysis (vary $\Delta R$, amplitude ratios 0/10/20 dB) with Monte Carlo random phases and fixed SNR.
% \end{itemize}
% \note{Keep intro short: 1 page max. Add 3--6 key references.}
\Ac{ISAC} aims to reuse mobile communication waveforms for environmental sensing while preserving their communication functionality. In this context, multiband sensing has recently emerged as a promising solution for OFDM-based \ac{ISAC}, since disjoint sensing bands can enlarge the effective frequency aperture and improve delay estimation when a wide contiguous sensing bandwidth is unavailable \cite{Wan24b}. This is particularly relevant for mobile systems, where contiguous spectrum is not always available, whereas carrier aggregation and sparse resource allocation are already part of the communication framework \cite{Wei23, Liu23}.

Existing works have investigated carrier aggregation, multiband fusion, and non-contiguous \ac{OFDM} sensing from both system and signal-processing viewpoints \cite{Wei23, Liu24, Ham23}. However, many state-of-the-art approaches rely on compressed-sensing-based recovery to handle missing frequency resources or sparse sensing observations \cite{Wei23, Liu23, Bat23}. Although effective in sparse regimes, these methods typically entail higher computational complexity, parameter tuning, and sensitivity to modeling assumptions, which motivates the search for lower-complexity and more structured reconstruction methods for practical multiband configurations.%%Although effective in sparse regimes, these methods often entail higher computational complexity, parameter tuning, and sensitivity to modeling assumptions such as sparsity level and source number \cite{Liu23, Wan24b}. This motivates the search for lower-complexity and more structured reconstruction methods tailored to practical multiband configurations.

Motivated by this gap, this work considers a coherent multiband sensing configuration with two equal-sized subbands symmetrically placed within the same \ac{OFDM} channel band (i.e., intra-band sensing), while the central portion remains available for communication or other services. The two sensing subbands are assumed to be coherently processed and sufficiently close in frequency that a common target reflectivity model applies to both. Building on this setting, we propose a novel iterative frequency-domain gap-reconstruction method composed of a kick-start delay-domain equalization stage and an iterative apodization-based operator with data-consistency enforcement, jointly reconstructing the missing subcarriers and suppressing grating lobes. Numerical results in multi-target scenarios show that the performance of the proposed method remains close to the full-band reference for moderate gap sizes and degrades only for larger gaps because of residual grating lobes. Under the same conditions, it outperforms the compressed-sensing-based \ac{OMP} baseline at low \acp{SNR} and shows more favorable complexity scaling as the number of targets increases.

In this paper, uppercase and lowercase bold letters denote matrices and vectors, respectively. The operators $(\cdot)^\mathsf{T}$, $(\cdot)^\mathsf{H}$, and $\|\cdot\|$ denote transpose, conjugate transpose, and Euclidean norm, respectively; $\mathbf{I}_n$ is the $n\times n$ identity matrix; $\mathbb{E}\{\cdot\}$ denotes expectation; and $\mathbf{x}\sim\mathcal{CN}(\mathbf{0},\boldsymbol{\Sigma})$ denotes a zero-mean circularly symmetric complex Gaussian vector with covariance $\boldsymbol{\Sigma}$. For a vector $\mathbf{x}$ indexed over $\mathcal K$ and a subset $\tilde{\mathcal K}\subset\mathcal K$, the selected subvector is denoted by $\mathbf{x}[\tilde{\mathcal K}]$. Moreover, $|\cdot|$ and $\angle\cdot$ denote the magnitude and argument of a scalar, respectively, $\odot$ denotes the Hadamard product, and $\operatorname{sgn}(\cdot)$ denotes the sign function.

The remainder of the paper is organized as follows: Section~\ref{sec:sys_mod} presents the system model; Section~\ref{sec:rec_method} describes the proposed iterative reconstruction method, the \ac{OMP} benchmark, and the associated complexity analysis; Section~\ref{sec:num_res} reports the numerical results; and Section~\ref{sec:conclusion} concludes the paper.
%The remainder of the paper is organized as follows: Section~\ref{sec:sys_mod} presents the system model; Section~\ref{sec:rec_method} describes the proposed iterative reconstruction method, the \ac{OMP} benchmark, and the associated complexity analysis; Section~\ref{sec:num_res} reports the numerical results; and Section~\ref{sec:conclusion} concludes the paper.
% ============================================================
\section{System Model}\label{sec:sys_mod}

\subsection{OFDM sensing observation model}
\label{sec:in_out_rel}
% ============================================================
%We consider a monostatic \ac{MIMO} \ac{OFDM}-based \ac{ISAC} system composed of a \ac{Tx} and a \ac{Rx} that are co-located and equipped with $N_\mathrm{T}$ and $N_\mathrm{R}$ antennas, respectively. Simultaneous transmission and reception are enabled by a full-duplex architecture that uses analog and digital self-interference cancellation techniques \cite{Barneto}. The system operates over a wireless channel with a carrier frequency $f_\mathrm{c}\gg B_\mathrm{max}$, where $B_\mathrm{max} = K \Delta f$ is the maximum bandwidth so that the narrowband array assumption holds \cite{van2002optimum}. Here, $K$ is the total number of active subcarriers indexed by the global index $k=0,\dots,K-1$, and $\Delta f = 1/T$ is the subcarrier spacing, with $T$  representing the \ac{OFDM} symbol duration excluding the cyclic prefix. The total \ac{OFDM} symbol duration is $T_\mathrm{s} = T + T_\mathrm{g}$, where $T_\mathrm{g}$ is the guard interval, used to prevent \ac{ISI}. The $k$th subcarrier frequency is given by $f_k = f_c + \left(k-\frac{K}{2}\right)\Delta f$.
%\end{equation}
%%%%%%%
We consider a monostatic OFDM-based \ac{ISAC} system equipped with a full-duplex architecture for simultaneous transmission and reception, with $N_\mathrm{T}$ transmit antennas and $N_\mathrm{R}$ receive antennas. The system operates over a wireless channel with carrier frequency $f_\mathrm{c}\gg B_\mathrm{max}$, where $B_\mathrm{max}=K\Delta f$ is the maximum bandwidth, so that the narrowband array assumption holds \cite{van2002optimum}. Here, $K$ is the total number of active subcarriers indexed by the global index $k=0,\dots,K-1$, and $\Delta f=1/T$ is the subcarrier spacing, with $T$ the \ac{OFDM} symbol duration excluding the cyclic prefix. The total symbol duration is $T_\mathrm{s}=T+T_\mathrm{g}$, where $T_\mathrm{g}$ is the guard interval, and the $k$th subcarrier frequency is $f_k = f_c + \left(k-\frac{K}{2}\right)\Delta f$. Sensing is performed over disjoint groups of contiguous subcarriers, hereinafter referred to as subbands. Each sensing subband $i$ occupies $K_i<K$ subcarriers and starts at index $\kappa_i$ on the full grid. The $n$th subcarrier of subband $i$, with local index $n=0,\dots,K_i-1$, corresponds to the global index $k=\kappa_i+n$.

% Under negligible \ac{ISI} and \ac{ICI}, the frequency-domain \ac{MIMO} sensing observation at the $n$th subcarrier of subband $i$ and OFDM symbol $m$ is
% %----------------
% \begin{equation}
%     \mathbf y_i[n,m]=\mathbf H_i[n,m]\mathbf x_i[n,m]+\mathbf n_i[n,m]
%     \label{eq:in-out-rel}
% \end{equation}
% %----------------
% where $\mathbf y_i[n,m]\in\mathbb C^{N_\mathrm R\times 1}$ is the received symbol vector, $\mathbf x_i[n,m]=\mathbf w_\mathrm T(\theta_\mathrm s)x_i[n,m]$ is the transmitted signal beamformed in the sensing direction $\theta_\mathrm{s}$, and $\mathbf n_i[n,m]\sim\mathcal{CN}(\mathbf 0,\sigma_n^2\mathbf I_{N_\mathrm R})$ is the noise term. %Residual self-interference is assumed to be negligible and absorbed into the noise term. 
% The transmit beamforming vector is normalized such that $\|\mathbf w_\mathrm T(\theta_\mathrm s)\|^2=P_\mathrm{avg}$, where $P_\mathrm{avg}$ denotes the average transmit power per subcarrier, and the transmitted symbols satisfy $\mathbb E\{|x_i[n,m]|^2\}=1$.
%  %The transmit beamforming vector $\mathbf w_\mathrm T(\theta_\mathrm s)$ is normalized such that $\|\mathbf w_\mathrm T(\theta_\mathrm s)\|^2=P_\mathrm{avg}$, where $P_\mathrm{avg}$ denotes the average transmit power per subcarrier. Moreover, the transmitted symbols $x_i[n,m]$ are drawn from a complex modulation alphabet whose elements satisfy $\mathbb{E}\{|x_i[n,m]|^2\}=1$.

In the presence of $N_\mathrm{s}$ scatterers, the frequency-domain $N_\mathrm{R}\times N_\mathrm{T}$ round-trip channel matrix at the $n$th subcarrier of subband $i$ and \ac{OFDM} symbol $m$ is modeled as
%%%%%%%%%%%%%%%
\begin{equation}
\label{eq:channel-matrix_ch2}
    \mathbf{H}_i[n,m]
    %& \triangleq \mathbf{H}[\kappa_i+n,m] \\
    = \sum_{l = 0}^{N_\mathrm{s}-1}
    \alpha_l e^{\imath 2\pi \left(m T_\mathrm{s} f_{\mathrm{D},l}-(\kappa_i+n)\Delta f \tau_l\right)}
    \mathbf{b}(\theta_l)\mathbf{a}^\mathsf{H}(\theta_l) %\nonumber
\end{equation}
%%%%%%%%%%%%%%
%where $\alpha_l$, $f_{\mathrm{D},l}$, $\theta_l$, and $\tau_l=2 r_l/c$ denote the complex channel coefficient, Doppler shift, target angle, and round-trip delay of the $l$th reflector, respectively.\footnote{In the monostatic case, \ac{AoA} and \ac{AoD} are treated as identical} The complex channel coefficient, accounting for round-trip attenuation and constant phase terms due to $f_c$ and frequency-grid centering, is assumed approximately invariant across the subcarriers of the considered OFDM channel band. Moreover, $r_l$ is the distance (range) between the $l$th target and the \ac{ISAC} transceiver, and $c$ is the speed of light. 
where $\alpha_l$, $f_{\mathrm D,l}$, $\theta_l$, and $\tau_l=\frac{2R_l}{c}$ denote the complex channel coefficient, Doppler shift, target angle, and round-trip delay of the $l$th reflector, respectively, with $R_l$ the target range and $c$ the speed of light. In the monostatic case, the angle of arrival and angle of departure coincide. The coefficient $\alpha_l$ accounts for round-trip attenuation and constant phase terms due to $f_c$, and is assumed approximately invariant across the considered \ac{OFDM} channel band. Moreover, $\mathbf a(\theta_l)$ and $\mathbf b(\theta_l)$ denote the transmit and receive steering vectors \cite{van2002optimum}.
%Lastly, $\mathbf{a}(\theta_l) \in \mathbb{C}^{N_\mathrm{T} \times 1}$ and $\mathbf{b}(\theta_l) \in \mathbb{C}^{N_\mathrm{R} \times 1}$ are the transmit and receive array response vectors, respectively, related to target $l$. Assuming the use of \acp{ULA} with inter-element spacing $\lambda/2$, with $\lambda = c/f_\mathrm{c}$ the carrier wavelength and taking the center of the array as a reference, the latter are defined as \cite{van2002optimum}
%Assuming \acp{ULA} with inter-element spacing $c/(2f_\mathrm{c})$, $\mathbf a(\theta_l)$ and $\mathbf b(\theta_l)$ denote the corresponding transmit and receive steering vectors \cite{van2002optimum}, with $\mathbf{a}(\theta_l) =\left [e^{-\imath \pi\frac{(N_\mathrm{T}-1)}{2} \sin(\theta_l)},\dots, e^{\imath \pi \frac{(N_\mathrm{T}-1)}{2}  \sin(\theta_l)} \right]^\mathsf{T}$ and  $\mathbf{b}(\theta_l) =\left [e^{-\imath \pi\frac{(N_\mathrm{R}-1)}{2} \sin(\theta_l)},\dots, e^{\imath \pi \frac{(N_\mathrm{R}-1)}{2}  \sin(\theta_l)} \right]^\mathsf{T}$.
% %%%%%%%%%%%%%%
% \begin{align}
%     \label{eq:steering_vectors_ch2}
%     \mathbf{a}(\theta_l) & =\left [e^{-\imath \pi\frac{(N_\mathrm{T}-1)}{2} \sin(\theta_l)},\dots, e^{\imath \pi \frac{(N_\mathrm{T}-1)}{2}  \sin(\theta_l)} \right]^\mathsf{T} \\
%     \mathbf{b}(\theta_l) & =\left [e^{-\imath \pi\frac{(N_\mathrm{R}-1)}{2} \sin(\theta_l)},\dots, e^{\imath \pi \frac{(N_\mathrm{R}-1)}{2}  \sin(\theta_l)} \right]^\mathsf{T}. \nonumber
% \end{align}
% %%%%%%%%%%%%%%

Under negligible \ac{ISI} and \ac{ICI}, the frequency-domain \ac{MIMO} sensing observation at the $n$th subcarrier of subband $i$ and \ac{OFDM} symbol $m$ is
\begin{equation}
\mathbf y_i[n,m]=\mathbf H_i[n,m]\mathbf x_i[n,m]+\mathbf n_i[n,m]
\label{eq:in-out-rel}
\end{equation}
where $\mathbf y_i[n,m]\in\mathbb C^{N_\mathrm R\times 1}$ is the received symbol vector, $\mathbf x_i[n,m]=\mathbf w_\mathrm T(\theta_\mathrm s)x_i[n,m]$ is the transmitted signal beamformed toward $\theta_\mathrm s$, and $\mathbf n_i[n,m]\sim\mathcal{CN}(\mathbf 0,\sigma_n^2\mathbf I_{N_\mathrm R})$ is the noise term. The transmit beamforming vector satisfies $\|\mathbf w_\mathrm T(\theta_\mathrm s)\|^2=P_\mathrm{avg}$, where $P_\mathrm{avg}$ denotes the average transmit power per subcarrier, and the transmitted symbols satisfy $\mathbb E\{|x_i[n,m]|^2\}=1$.

%This work focuses on the impact of considering disjoint sensing bands on the formation of delay profiles (i.e., range profiles). For this reason, we focus hereinafter on a scenario where beamforming is performed in a given sensing direction $\theta_\mathrm{s}$ at both the transmitter and the receiver. Moreover, the proposed reconstruction is applied to the frequency-domain samples of a representative OFDM symbol, so the symbol index $m$ and the related Doppler shift terms are omitted. Under these hypotheses, by plugging \eqref{eq:channel-matrix_ch2} in \eqref{eq:in-out-rel} and applying reciprocal filtering to remove the dependence on the transmitted symbols \cite{Kes:J25}, we obtain the sensing received symbol at subcarrier $n$ of subband $i$, as follows\cite{PucPaoGio:J22}
Since this work focuses on delay-profile formation, we consider a representative \ac{OFDM} symbol and omit the symbol index $m$ and the Doppler dependence hereinafter. Moreover, we focus on a scenario where beamforming is performed in a given sensing direction $\theta_\mathrm{s}$ at both the transmitter and the receiver. Under these assumptions, by combining \eqref{eq:channel-matrix_ch2} and \eqref{eq:in-out-rel} and applying reciprocal filtering \cite{Kes:J25}, the sensing received symbol at subcarrier $n$ of subband $i$ is
%%%%%%%%%%%%%%%%%%
\begin{equation}
    \label{eq:rx-sig}
    r_i[n]  \triangleq r[\kappa_i+n] %&= \frac{\mathbf{w}^\mathsf{H}_\mathrm{R}(\theta_\mathrm{s}) \mathbf{y}_i[n]}{x_i[n]}\nonumber\\
    %& 
    = \sum_{l = 0}^{N_\mathrm{s}-1}
    \gamma_l e^{-\imath 2\pi (\kappa_i+n)\Delta f \tau_l}
    + \nu_i[n]
\end{equation}
%%%%%%%%%%%%%%
where $\mathbf{w}_\mathrm{R}$ is the receive beamforming vector, normalized such that $\|\mathbf{w}_\mathrm{R}(\theta_\mathrm{s})\|^2=1$, $\gamma_l \triangleq \alpha_l\,\mathbf{w}^\mathsf{H}_\mathrm{R}(\theta_\mathrm{s}) \mathbf{b}(\theta_l)\mathbf{a}^\mathsf{H}(\theta_l) \mathbf{w}_\mathrm{T}(\theta_\mathrm{s})$ is the effective post-beamforming complex coefficient, and $\nu_i[n] = \tfrac{\mathbf{w}^\mathsf{H}_\mathrm{R}\mathbf{n}_i[n]}{x_i[n]}$ is the effective noise term.  If constant-envelope symbols are used for sensing, e.g., \ac{PSK}, then $\nu_i[n]$ remains zero-mean complex Gaussian with variance $\sigma_\nu^2=\sigma^2_n$. The corresponding post-beamforming \ac{SNR} of target $l$ is defined as%\footnote{Alternatively, the \ac{SNR} can be defined for each receive antenna as in \cite{PucPaoGio:J22}. In that case, the receive beamforming gain is implicitly accounted for as a processing gain, resulting in a leftward shift in the system performance.}
%%%%%%%%%%%%%%%%
\begin{equation}
    \mathrm{SNR}_l = \frac{|\gamma_l|^2}{\sigma^2_\nu}.
    \label{eq:SNR}
\end{equation}
%%%%%%%%%%%%%%%
\subsection{Coherent dual-band OFDM sensing}
As mentioned, we consider a symmetric dual-band sensing configuration, where two sensing subbands of equal size are located at the edges of a total bandwidth comprising $K$ subcarriers. %, as schematically shown in Fig.~\ref{fig:dual_band_sensing}.
Let $\rho_\mathrm{g}\in[0,1)$ denote the fraction of the total bandwidth occupied by the intermediate gap, so that $\rho_\mathrm{g}=K_\mathrm{g}/K$, where $K_\mathrm{g}$ denotes the number of gap subcarriers. The number of subcarriers in each sensing subband is then
%%%%%
\begin{equation}
    K_\mathrm{s}=\frac{K-K_\mathrm{g}}{2}
    \label{eq:Ks}
\end{equation}
%%%%%%%
where $K_\mathrm{g}$ is chosen as the integer closest to $\rho_\mathrm{g}K$ such that $K-K_\mathrm{g}$ is even. The $K_\mathrm{g}$ intermediate subcarriers are not used for sensing and are therefore treated as missing frequency-domain samples.
%The $K_\mathrm{g}$ intermediate subcarriers are not used for sensing and are therefore modeled as missing samples in the frequency domain, although they belong to the same overall bandwidth. 
From \eqref{eq:rx-sig}, we define the vector of receive frequency-domain samples at each subband $i$ as 
\[
\mathbf r_i = [r_i[0],\dots,r_i[K_\mathrm{s}-1]]^\mathsf T,
\qquad i\in\{1,2\}
\]
%where the elements of $\mathbf{r}_1$ are computed for $\kappa_1 = 0$ while for $\mathbf{r}_2$ we set $\kappa_2 = K_\mathrm{s} + K_\mathrm{g}$. The two sequences are then aggregated into the composite vector $\mathbf r_\mathrm{CA}\in\mathbb C^{K\times 1}$ as
where $\kappa_1 = 0$ for $\mathbf{r}_1$ and $\kappa_2 = K_\mathrm{s} + K_\mathrm{g}$ for $\mathbf{r}_2$. The resulting $K\times 1$ composite observation is
%%%%%%%%%%%%%%%%
\begin{equation}
    \mathbf r_\mathrm{CA}
    =
    [\mathbf r_1^\mathsf T,\mathbf 0_{1\times K_\mathrm{g}},\mathbf r_2^\mathsf T]^\mathsf T.
    \label{eq:rx-signal-CA}
\end{equation}
%%%%%%%%%%%%%%%%
%The composite multiband observation in \eqref{eq:rx-signal-CA} can be interpreted as a full-band frequency-domain signal, i.e., with $K$ continuous subcarriers used for sensing, $\mathbf r_\mathrm{full}\in\mathbb C^{K\times 1}$ masked by the binary window $\mathbf w_\mathrm{CA}$, i.e.,
Equivalently, $\mathbf r_\mathrm{CA}$ can be viewed as a full-band signal $\mathbf r_\mathrm{full}\in\mathbb C^{K\times1}$ masked by the binary window $\mathbf w_\mathrm{CA}$, i.e.,
\begin{equation}
    \mathbf r_\mathrm{CA} = \mathbf r_\mathrm{full}\odot \mathbf w_\mathrm{CA}
    \label{eq:r_ca_masking}
\end{equation}
where $\mathbf w_\mathrm{CA}$ is defined entrywise as
%%%%%%%%%%%%%%%
\begin{equation}
    w_\mathrm{CA}[k] =
    \begin{cases}
        1, & k \in \mathcal K_\mathrm{CA},\\
        0, & \text{otherwise}
    \end{cases}
    \qquad k=0,\dots,K-1
    \label{eq:fft_window}
\end{equation}
%%%%%%%%%%%%%
with $\mathcal{K}_\mathrm{CA} = \{0,\dots,K_\mathrm{s}-1\} \; \cup \; \{K_\mathrm{s}+K_\mathrm{g}, \dots, K - 1\}$.
% %%%%%%%%%
% \begin{figure}[t]
%     \centering
%     \includegraphics[width=0.8\columnwidth]{Fig/multiband_new.pdf}
%     \caption{Illustration of dual-band sensing over a frequency grid of size $K=2K_\mathrm{s}+K_\mathrm{g}$.}
%     \label{fig:dual_band_sensing}
% \end{figure}
% %%%%%%%%%%%%%%
\subsection{Delay-response sidelobe analysis}
The vector $\mathbf r_\mathrm{CA}$ is used to compute the normalized delay response as \cite{braun2010maximum}\footnote{When evaluated on the discrete delay grid $\tau_q= q \Delta \tau = q/(K\Delta f)$, $q=0,\dots,K-1$, \eqref{eq:period} coincides with the $K$-point \ac{IDFT} of $\mathbf r_\mathrm{CA}$ up to normalization.}
\begin{equation}
    \mathcal P(\tau)
    \triangleq
    \frac{1}{\mathrm{card}(\mathcal K_\mathrm{CA})}
    \sum_{k=0}^{K-1} r_\mathrm{CA}[k] e^{\imath 2\pi k \Delta f \tau}.
    \label{eq:period}
\end{equation}
To characterize the grating-lobe structure introduced by the spectral gap, consider a noise-free single-target scenario with $N_\mathrm{s}=1$, $\gamma_0=1$, and target delay $\tau_0$. From \eqref{eq:rx-sig}, in the corresponding full-band noiseless single-target case, the frequency-domain samples are
\begin{equation}
    r_\mathrm{full}[k] = e^{-\imath 2\pi k \Delta f \tau_0},
    \qquad k=0,\dots,K-1
\end{equation}
and from \eqref{eq:r_ca_masking} we have $r_\mathrm{CA}[k] = w_\mathrm{CA}[k] e^{-\imath 2\pi k \Delta f \tau_0}$.
Substituting into \eqref{eq:period} yields
%%%%%%%%
\begin{align}
    \label{eq:period_exp}
    \bar{\mathcal P}(\tau)
    &= \frac{1}{2K_\mathrm{s}}
    \sum_{k\in\mathcal K_\mathrm{CA}}
    e^{\imath 2\pi k \Delta f (\tau-\tau_0)}\\
    %&= \frac{1}{2K_\mathrm{s}}
    %\left(
   % \sum_{k=0}^{K_\mathrm{s}-1} e^{\imath 2\pi k \Delta f (\tau-\tau_0)}
  %  +
  %  \sum_{k=K_\mathrm{s}+K_\mathrm{g}}^{K-1}
  %  e^{\imath 2\pi k \Delta f (\tau-\tau_0)}
%    \right)
    %\nonumber\\
    &= \frac{1}{2K_\mathrm{s}}
    \left(
    1 + e^{\imath 2\pi (K_\mathrm{s}+K_\mathrm{g})\Delta f(\tau-\tau_0)}
    \right)
    \frac{1-e^{\imath 2\pi K_\mathrm{s}\Delta f(\tau-\tau_0)}}
         {1-e^{\imath 2\pi \Delta f(\tau-\tau_0)}}
    \nonumber\\
    &= \frac{1}{K_\mathrm{s}}
    e^{\imath \pi (K-1)\Delta f(\tau-\tau_0)}
    \cos\!\bigl(\pi (K_\mathrm{s}+K_\mathrm{g})\Delta f(\tau-\tau_0)\bigr) \nonumber \\
    & \quad \times
    \frac{\sin\!\bigl(\pi K_\mathrm{s}\Delta f(\tau-\tau_0)\bigr)}
         {\sin\!\bigl(\pi \Delta f(\tau-\tau_0)\bigr)}. \nonumber 
\end{align}
%%%%%%
From \eqref{eq:period_exp}, it can be noticed that the delay response is the product of a Dirichlet-kernel term associated with $K_\mathrm{s}$ contiguous subcarriers and a cosine modulation induced by the separation between the two sensing subbands. %The Dirichlet term determines the overall envelope of the response, while the cosine term introduces an oscillatory fine structure that generates grating lobes. 
As $K_\mathrm{g}$ increases for fixed $K$, $K_\mathrm{s}$ decreases, so the Dirichlet envelope broadens. At the same time, the cosine modulation oscillates faster with $\tau$, producing more pronounced secondary peaks. Their interplay results in an increasingly structured grating-lobe pattern in the delay response. As a sanity check, for $K_\mathrm{g}=0$, by setting $a=\pi K_\mathrm{s}\Delta f (\tau-\tau_0)$ and using $2\sin(a)\cos(a)=\sin(2a)$, \eqref{eq:period_exp} reduces to the normalized full-band Dirichlet response associated with $K=2K_\mathrm{s}$ contiguous subcarriers.
%%%%
% ============================================================
\begin{figure}[t]
    \centering
    \includegraphics[width=0.93\linewidth]{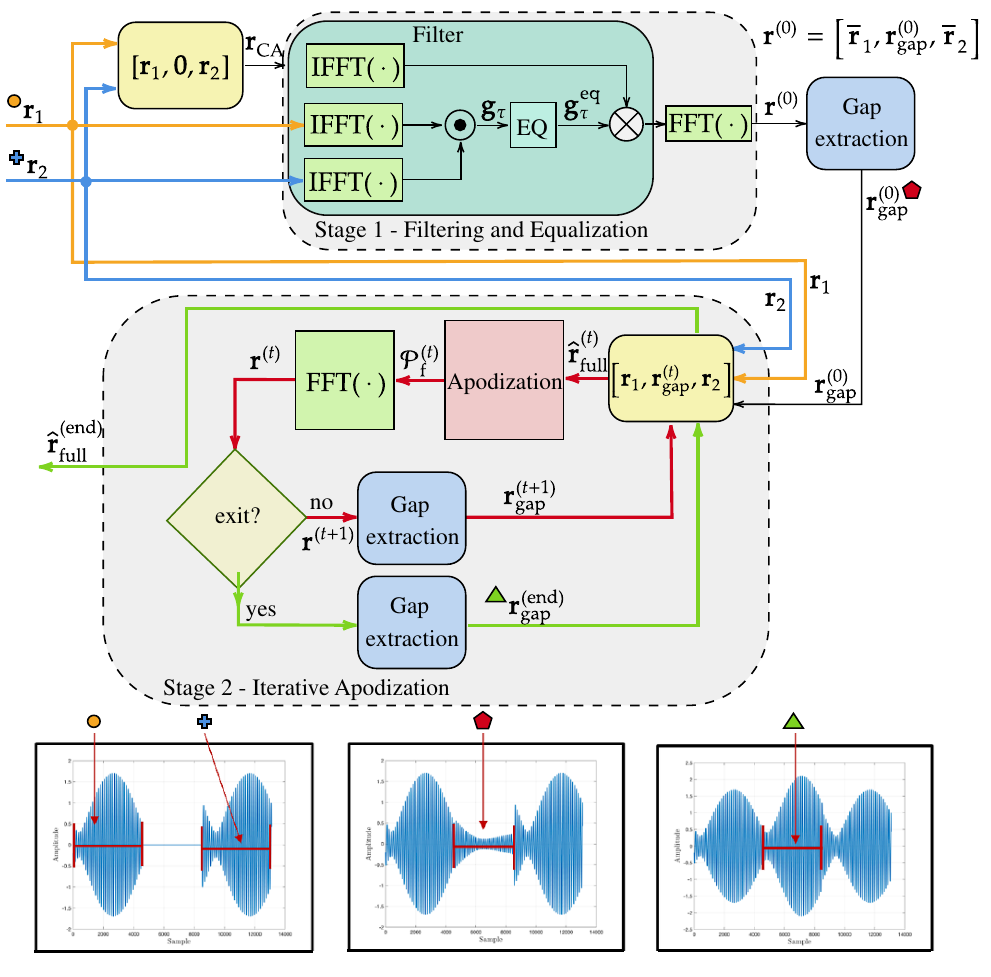}
    \caption{Block diagram of the proposed iterative reconstruction method.}
    \label{fig:iter_method}
\end{figure}
%%%
\section{Proposed Reconstruction Method}
\label{sec:rec_method}
%%%
To jointly reconstruct the missing subcarriers and suppress grating lobes, we propose a two-stage iterative method, schematically shown in Fig.~\ref{fig:iter_method}. The first stage performs a kick-start delay-domain equalization, which builds an equalization mask from the two observed subbands to mitigate the distortion induced by the spectral gap and provide an improved initialization. The second stage iteratively refines the reconstruction of the missing subcarriers through an apodization-based operator, while enforcing data consistency by keeping the observed samples fixed. The two stages are detailed in Sections~\ref{sec:equalization} and \ref{sec:iter_complex_apo}, respectively.
%To jointly reconstruct the missing subcarriers and suppress grating lobes, we propose an iterative method composed of two main stages: i) an initial delay-domain equalization, and ii) an iterative nonlinear complex apodization with data-consistency enforcement on the observed subcarriers. A schematic overview is shown in Fig.~\ref{fig:iter_method}.

%The first stage builds a delay-domain equalization mask from the two active subbands to compensate for the distortion induced by the spectral gap and provide an improved initialization for the subsequent reconstruction. Applied to $\mathbf{r}_\mathrm{CA}$ prior to delay-profile evaluation, this step mitigates grating lobes. The corresponding design is detailed in Section~\ref{sec:equalization}.

%The second stage iteratively reconstructs the missing subcarriers by nonlinear complex apodization, while the observed samples are kept fixed to enforce data consistency. Further details are provided in Section~\ref{sec:iter_complex_apo}.

\subsection{Kick-start delay-domain equalization}
\label{sec:equalization}

To initialize the reconstruction, we adopt a gap-aware equalization kernel from the two observed sensing subbands. Specifically, we define the auxiliary frequency-domain sequence
%\begin{equation}
    $\mathbf g_\mathrm{f} = \mathbf r_1 \ast \mathbf r_2$,
    %\label{eq:conv_filter}
%\end{equation}
whose entries are given by
\begin{equation}
    g_\mathrm{f}[n] = \sum_{k=0}^{K_\mathrm{s}-1} r_1[k]\, r_2[n-k],
    \qquad n=0,\dots,2K_\mathrm{s}-2
    \label{eq:conv_filter_expanded}
\end{equation}
where out-of-range samples of $r_2[\cdot]$ are set to zero. The sequence $\mathbf g_\mathrm{f}\in\mathbb C^{(2K_\mathrm{s}-1)\times 1}$ is used as an auxiliary kernel to enforce the delay-domain equalization through its linear convolution with the composite multiband observation, i.e., $\mathbf r_\mathrm{CA} \ast \mathbf g_\mathrm{f}$. 
%can be interpreted as an auxiliary kernel built from the two observed subbands. Its linear convolution with the composite multiband observation, i.e., $\mathbf r_\mathrm{CA} \ast \mathbf g_\mathrm{f}$, motivates the delay-domain equalization procedure developed below.

Let $K_\mathrm{p,reb}\geq K+2K_\mathrm{s}-2$ be a power of two, and let $\tilde{\mathbf r}_1$, $\tilde{\mathbf r}_2$, and $\tilde{\mathbf r}_\mathrm{CA}$ denote the zero-padded versions of $\mathbf r_1$, $\mathbf r_2$, and $\mathbf r_\mathrm{CA}$, respectively, all of size $K_\mathrm{p,reb}\times 1$. We define the prototype delay-domain response as
\begin{equation}
    \mathbf g_\tau
    =
    \mathrm{IFFT}_{K_\mathrm{p,reb}}(\tilde{\mathbf r}_1)
    \odot
    \mathrm{IFFT}_{K_\mathrm{p,reb}}(\tilde{\mathbf r}_2)
    \label{eq:G_prototype}
\end{equation}
where $\mathrm{IFFT}_{K_\mathrm{p,reb}}(\cdot)$ denotes the $K_\mathrm{p,reb}$-point \ac{IFFT}. %The vector $\mathbf g_\tau\in\mathbb C^{K_\mathrm{p,reb}\times 1}$ captures the prototype delay-domain structure induced by the two observed subbands.
In multi-target scenarios, the delay-domain power response $\mathbf h = |\mathbf g_\tau|^2$ is typically dominated by the strongest target. 
%As a result, directly using $\mathbf g_\tau$ may lead to an excessive attenuation of weaker targets. To alleviate this effect, Algorithm~\ref{alg:adaptive_equalizer}, which operates on the normalized delay-domain power response in dB, i.e., $\mathbf h_{\mathrm{dB}} = 10\log_{10}\!\left(\frac{\mathbf h}{\max(\mathbf h)}\right)$, applies adaptive tapered equalization to the normalized delay-domain power response before constructing the final complex mask.
To alleviate the resulting attenuation of weaker targets, Algorithm~\ref{alg:adaptive_equalizer} applies adaptive tapered equalization to the normalized delay-domain power response in dB,  i.e., $\mathbf h_{\mathrm{dB}} = 10\log_{10}\!\left(\frac{\mathbf h}{\max(\mathbf h)}\right)$ before constructing the final mask.
%As a result, directly using $\mathbf g_\tau$ may lead to an excessive attenuation of weaker targets. To alleviate this effect, Algorithm~\ref{alg:adaptive_equalizer}, which operates on the normalized delay-domain power response in dB, i.e., $\mathbf h_{\mathrm{dB}} = 10\log_{10}\!\left(\frac{\mathbf h}{\max(\mathbf h)}\right)$, applies adaptive tapered equalization to the normalized delay-domain power response before constructing the final complex mask.

First, the dominant peaks of $\mathbf h_{\mathrm{dB}}$ above a threshold $\eta_\mathrm{dB}$ are identified and sorted in descending order of amplitude. If multiple peaks are present, the strongest one is kept unchanged, whereas up to the next $J_{\max}-1$ secondary peaks are compensated through Gaussian-shaped gain masks centered at their delay locations. The spread of each mask is selected from the corresponding peak width through a width-to-standard-deviation conversion factor $c_w$, and overlapping masks are combined by pointwise maximization. %The spread of each mask is selected according to the corresponding peak width, and overlapping masks are combined by pointwise maximization. 
%The resulting equalized response in dB is denoted by $\mathbf h_{\mathrm{eq,dB}}$, with corresponding linear-scale version $\mathbf h_\mathrm{eq} = 10^{\mathbf h_{\mathrm{eq,dB}}/10}$.
%The corresponding equalized complex delay-domain mask is then obtained by preserving the phase of $\mathbf g_\tau$, as follows
Denoting the resulting equalized response in dB by $\mathbf h_{\mathrm{eq,dB}}$, its linear-scale counterpart is $\mathbf h_\mathrm{eq}=10^{\mathbf h_{\mathrm{eq,dB}}/10}$, and the corresponding complex equalization mask is
\begin{equation}
    \mathbf g_\tau^\mathrm{eq}
    =
    \sqrt{\mathbf h_\mathrm{eq}}
    \odot
    e^{\imath\angle \mathbf g_\tau}.
    \label{eq:G_eq}
\end{equation}

The Stage-1 equalized initialization is obtained by applying the normalized complex mask $\mathbf g_\tau^\mathrm{eq}$ to the delay-domain representation of the zero-padded multiband observation, namely
\begin{equation}
    \tilde{\mathbf r}^{(0)}
    =
    \mathrm{FFT}_{K_\mathrm{p,reb}}
    \Bigl(
    \mathbf g_\tau^\mathrm{eq}
    \odot
    \mathrm{IFFT}_{K_\mathrm{p,reb}}(\tilde{\mathbf r}_\mathrm{CA})
    \Bigr)
    \label{eq:fft_conv}
\end{equation}
where $\mathrm{FFT}_{K_\mathrm{p,reb}}(\cdot)$ denotes the $K_\mathrm{p,reb}$-point \ac{FFT}. Unitary FFT/IFFT conventions are adopted throughout, so that both the $K_\mathrm{p,reb}$-point FFT and IFFT include the normalization factor $1/\sqrt{K_\mathrm{p,reb}}$. %\footnote{Unitary FFT/IFFT conventions are adopted throughout, so that both the $K_\mathrm{p,reb}$-point FFT and IFFT include the normalization factor $1/\sqrt{K_\mathrm{p,reb}}$.}.
Since $\mathbf g_\tau^\mathrm{eq}$ is normalized to unit peak gain, \eqref{eq:fft_conv} is used here as a delay-domain shaping/equalization step to provide an improved initialization, rather than as an exact convolution-preserving implementation.

The initial full-band estimate is then obtained by retaining the first $K$ entries of $\tilde{\mathbf r}^{(0)}$, i.e.,
\begin{equation}
    \mathbf r^{(0)}
    =
    \bigl[\tilde r^{(0)}[0],\ldots,\tilde r^{(0)}[K-1]\bigr]^\mathsf T.
    \label{eq:r0_init}
\end{equation}
Since the observed subband samples $\mathbf r_1$ and $\mathbf r_2$ are already available, Stage~1 is used only to initialize the missing gap samples. Therefore, letting $\mathcal K_\mathrm{gap}=\{K_\mathrm{s},\dots,K_\mathrm{s}+K_\mathrm{g}-1\}$, the initialization passed to Stage~2 is
\begin{equation}
    \mathbf r_\mathrm{gap}^{(0)}
    =
    \mathbf r^{(0)}[\mathcal K_\mathrm{gap}].
   \label{eq:gap_init}
\end{equation}
%-------------------------------------%
\begin{algorithm}[t]
\caption{Adaptive tapered equalization}
\label{alg:adaptive_equalizer}
\begin{algorithmic}[1]
\Function{Equalize}{$\mathbf h_{\mathrm{dB}}, \eta_{\mathrm{dB}}, J_{\max}$}
    \State Detect and sort peaks $\{(p_j,\ell_j,w_j)\}_{j=1}^{N_{\mathrm{p}}}$ with $p_j>\eta_{\mathrm{dB}}$
    \If{$N_{\mathrm{p}}\leq 1$}
        \State \Return $\mathbf h_{\mathrm{dB}}$
    \EndIf
    \State $\Delta \mathbf h_{\mathrm{dB}} \gets \mathbf 0$
    \For{$j=2,\dots,\min(N_{\mathrm{p}},J_{\max})$}
        \State $\sigma_j \gets w_j/c_w$, \quad $\alpha_j \gets -h_{\mathrm{dB}}[\ell_j]$
        \For{each bin $\ell$ with $|\ell-\ell_j|\leq 4\sigma_j$}
            \State $\Delta h_{\mathrm{dB}}[\ell] \gets \max\!\left(\Delta h_{\mathrm{dB}}[\ell],\, \alpha_j e^{-(\ell-\ell_j)^2/(2\sigma_j^2)}\right)$
        \EndFor
    \EndFor
    \State \Return $\mathbf h_{\mathrm{eq,dB}} = \min(\mathbf h_{\mathrm{dB}} + \Delta \mathbf h_{\mathrm{dB}}, \mathbf 0)$
\EndFunction
\end{algorithmic}
\end{algorithm}
%===================================%
\subsection{Iterative apodization-based operator}
\label{sec:iter_complex_apo}
%===================================%
%Apodization originates from optics and refers to the suppression of diffraction sidelobes. In \ac{SAR}, it is widely used to reduce sidelobes in range/Doppler responses by weighting the data with suitable window functions \cite{Apodization:J95}. 
%In this work, we exploit the same principle to iteratively refine the reconstructed delay profile while enforcing consistency with the observed subcarriers.
Apodization is a principle originating from optics and is used in \ac{SAR} to reduce sidelobes in range/Doppler responses by weighting the data with suitable window functions \cite{Apodization:J95}.
In this work, the same principle is applied iteratively to refine the reconstructed delay profile while enforcing consistency with the observed subcarriers.

Following Fig.~\ref{fig:iter_method}, Stage~2 is initialized from the gap estimate obtained in Stage~1. The initial full-band estimate is therefore defined as
$%\begin{equation}
\hat{\mathbf r}_{\mathrm{full}}^{(0)}
=
[\mathbf r_1^\mathsf T,(\mathbf r_\mathrm{gap}^{(0)})^\mathsf T,\mathbf r_2^\mathsf T]^\mathsf T$, 
%\label{eq:rfull_init_stage2}
%\end{equation}
where $\mathbf r_\mathrm{gap}^{(0)}$ is given in \eqref{eq:gap_init}. At iteration $t$, the current full-band estimate is
\begin{equation}
\hat{\mathbf r}_{\mathrm{full}}^{(t)}
=
[\mathbf r_1^\mathsf T,(\mathbf r_\mathrm{gap}^{(t)})^\mathsf T,\mathbf r_2^\mathsf T]^\mathsf T.
\label{eq:rfull_t}
\end{equation}
\begin{figure*}[t]
    \centering
    \includegraphics[width=0.76\linewidth]{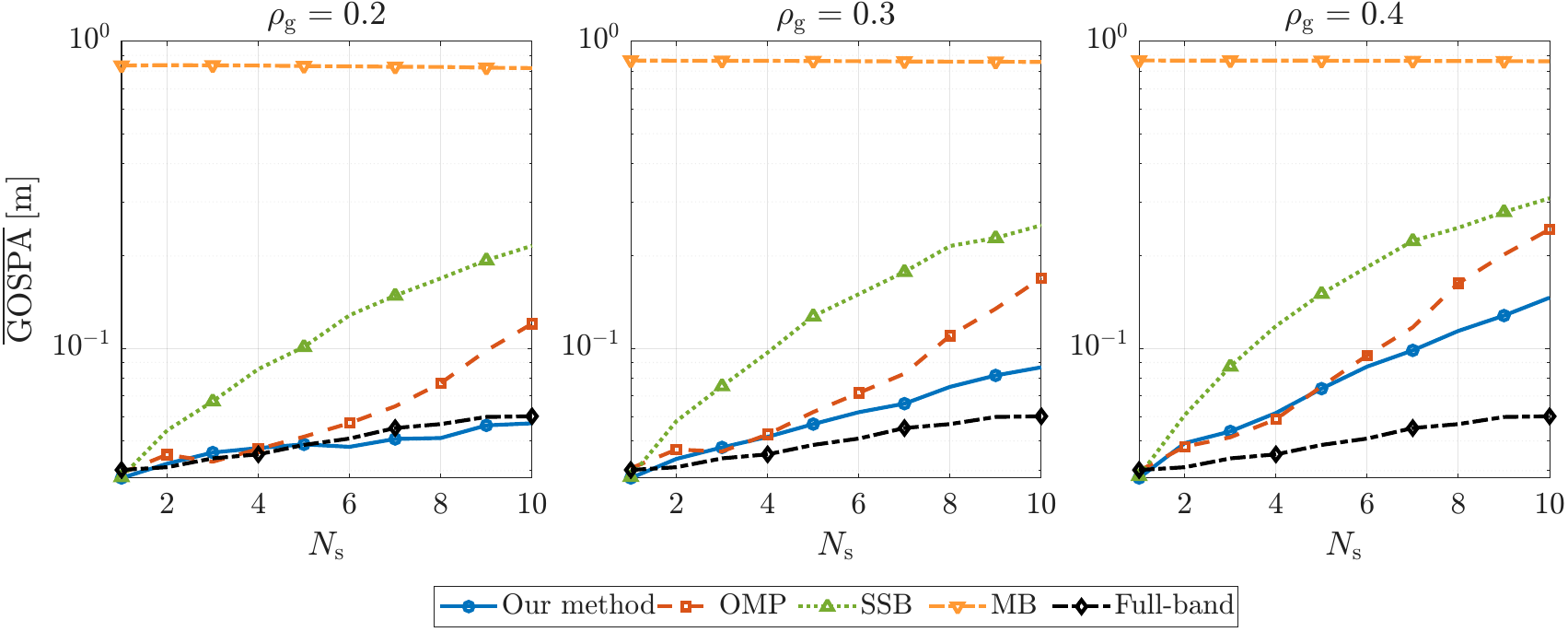}
    \caption{Mean GOSPA as a function of the number of targets $N_\mathrm{s}$ for $\mathrm{SNR}= -5\,$dB.}
    \label{fig:GOSPA_vs_Ntg}
\end{figure*}

Let $\mathbf w^{(j)} \in \mathbb R^{K\times 1}$ denote the $j$th apodization window, normalized to unit average power, for $j=1,\dots,N_\mathrm W$. The corresponding windowed frequency-domain signal is
\begin{equation}
\mathbf r^{(j,t)}
\triangleq
\mathbf w^{(j)} \odot \hat{\mathbf r}_{\mathrm{full}}^{(t)}.
\label{eq:win_apply}
\end{equation}
%Let $\tilde{\mathbf r}^{(j,t)}\in\mathbb C^{K_\mathrm p\times 1}$ denote the zero-padded version of $\mathbf r^{(j,t)}$, with $K_\mathrm p\ge K$. Since the goal is to identify and suppress grating lobes, $K_\mathrm p$ is typically chosen sufficiently larger than $K$ to obtain a finer delay grid. This enables more accurate resolution and comparison of the grating-lobe structures across the $N_\mathrm {W} $-apodized profiles, thereby facilitating their identification and suppression by the nonlinear selection rule. 
Let $\tilde{\mathbf r}^{(j,t)}\in\mathbb C^{K_\mathrm p\times 1}$ denote the zero-padded version of $\mathbf r^{(j,t)}$, with $K_\mathrm p\ge K$. To better resolve and compare grating-lobe structures across the $N_\mathrm W$-apodized profiles, $K_\mathrm p$ is chosen larger than $K$, yielding a finer delay grid.
The associated discrete delay response is then computed as
\begin{equation}
\boldsymbol{\mathcal P}^{(j,t)}
=
\mathrm{IFFT}_{K_\mathrm p}\!\left(\tilde{\mathbf r}^{(j,t)}\right)
\label{eq:Pc_m_iter}
\end{equation}
whose $q$th entry is denoted by
\[
\mathcal P^{(j,t)}[q]
=
\mathcal P_\mathrm I^{(j,t)}[q]
+
\imath\,\mathcal P_\mathrm Q^{(j,t)}[q],
\qquad q=0,\dots,K_\mathrm p-1.
\]

\paragraph*{Component-wise selection rule}
For each delay bin $q$, the real and imaginary parts of the $N_\mathrm W$ delay responses are processed separately. %Let $C^{(j,t)}[q]\in\left\{\mathcal P_\mathrm I^{(j,t)}[q],\,\mathcal P_\mathrm Q^{(j,t)}[q]\right\}$ denote one of the two components. 
Let $C^{(j,t)}[q]$ denote either $\mathcal P_\mathrm I^{(j,t)}[q]$ or $\mathcal P_\mathrm Q^{(j,t)}[q]$. The filtered component is obtained by enforcing sign consistency across the $N_\mathrm W$ windowed responses and, when satisfied, selecting the sample with minimum magnitude, i.e.,
\begin{equation}
C_\mathrm f^{(t)}[q]
=
\begin{cases}
C^{(j^\star_{C,q,t},t)}[q], &
\text{if }\left|\sum_{j=1}^{N_\mathrm W}\operatorname{sgn}\!\big(C^{(j,t)}[q]\big)\right|=N_\mathrm W,\\
0, & \text{otherwise}
\end{cases}
\label{eq:complex_apod_rule}
\end{equation}
with 
%\begin{equation}
$j^\star_{C,q,t}\in
\arg\min_{j\in\{1,\dots,N_\mathrm W\}}
\big|C^{(j,t)}[q]\big|.$
%\label{eq:argminC}
%\end{equation}

\noindent The filtered complex delay profile is then
\begin{equation}
\mathcal P_\mathrm f^{(t)}[q]
=
\mathcal P_{\mathrm I,\mathrm f}^{(t)}[q]
+
\imath\,\mathcal P_{\mathrm Q,\mathrm f}^{(t)}[q],
\qquad q=0,\dots,K_\mathrm p-1.
\label{eq:Pf_complex}
\end{equation}

In this work, $N_\mathrm W=3$ and the adopted windows are rectangular, Hamming, and Hann (tri-apodization) \cite{Apodization:J95}. Transforming back to the frequency domain yields
%Transforming the filtered delay profile back to the frequency domain yields
\begin{equation}
\tilde{\mathbf r}^{(t+1)}
=
\mathrm{FFT}_{K_\mathrm p}\!\left(\boldsymbol{\mathcal P}_\mathrm f^{(t)}\right)
\label{eq:fft_back_iter}
\end{equation}
from which the first $K$ entries are retained to obtain
\begin{equation}
\mathbf r^{(t+1)}
=
\bigl[\tilde r^{(t+1)}[0],\ldots,\tilde r^{(t+1)}[K-1]\bigr]^\mathsf T.
\label{eq:r_update_iter}
\end{equation}
Therefore, the updated gap estimate is
\begin{equation}
\mathbf r_\mathrm{gap}^{(t+1)}
=
\mathbf r^{(t+1)}[\mathcal K_\mathrm{gap}]
\label{eq:gap_update_iter}
\end{equation}
and the full-band vector for the next iteration is formed as
\begin{equation}
\hat{\mathbf r}_{\mathrm{full}}^{(t+1)}
=
[\mathbf r_1^\mathsf T,(\mathbf r_\mathrm{gap}^{(t+1)})^\mathsf T,\mathbf r_2^\mathsf T]^\mathsf T.
\label{eq:rfull_update_iter}
\end{equation}
Hence, only the missing gap samples are updated, while the observed subcarriers are kept fixed.

The above procedure is repeated for a prescribed number of iterations $T_{\max}$ until the final reconstructed signal $\hat{\mathbf r}^{(\mathrm{end})}_\mathrm{full}$ is obtained.\footnote{In this work, $T_{\max}$ is selected as a function of the spectral gap configuration based on a preliminary offline convergence analysis and then kept fixed throughout the simulations. A more complete convergence analysis, including adaptive stopping criteria, is left for future work.}
%In this work, $T_{\max}$ is selected as a function of the spectral gap configuration based on a preliminary offline convergence analysis and then kept fixed throughout the simulations. A more complete convergence analysis, including adaptive stopping criteria, is left for future work.}
Finally, $\hat{\mathbf r}^{(\mathrm{end})}_\mathrm{full}\in\mathbb C^{K\times1}$ is used to compute the complex delay response according to \eqref{eq:period}, which can then be mapped to range through the round-trip delay--range relationship in Section~\ref{sec:in_out_rel}. Equivalently, the complex range response can be efficiently evaluated as the \ac{IFFT} of the reconstructed signal \cite{PucPaoGio:J22}, while the range power profile adopted for estimation is obtained from its squared magnitude. When needed, zero padding with $K_\mathrm{p,est}\geq K$ provides a finer \ac{IFFT} range grid. The actual range resolution is $\Delta r_\mathrm{full}=\frac{c}{2K\Delta f}$, whereas the corresponding \ac{IFFT} range-grid spacing is $\Delta r_{\mathrm{grid}}=\frac{c}{2K_\mathrm{p,est}\Delta f}$.

\subsection{OMP benchmark and complexity analysis}
\label{sec:omp_complexity}

As a benchmark, we consider a conventional \ac{OMP}-based reconstruction in which, for the single-symbol case, the missing subcarriers are recovered through a sparse expansion over an oversampled delay dictionary. Owing to the harmonic structure of the atoms, the correlation step can be implemented efficiently through a $K_{\mathrm{p,omp}}$-point \ac{IFFT}, while the least-squares update is performed recursively via QR factorization \cite{eldar2012compressed}.

Let $L$ denote the number of atoms selected by \ac{OMP}, corresponding to the estimated number of targets, and let $N_\mathrm{on}=2K_\mathrm{s}$ be the number of observed subcarriers. The resulting complexity of the \ac{OMP} benchmark scales as
\begin{equation}
\mathcal O\!\left(
L K_{\mathrm{p,omp}}\log_2 K_{\mathrm{p,omp}}
+
N_{\mathrm{on}}L^2
+
L^3
\right)
\label{eq:omp_complexity}
\end{equation}
where the first term accounts for the repeated transform-based correlation search, whereas the remaining terms arise from the recursive QR/least-squares updates.

For the proposed method, the dominant cost is due to FFT/IFFT operations. Stage-1 requires three $K_\mathrm p$-point \acp{IFFT} and one $K_\mathrm p$-point \ac{FFT}, while the additional Gaussian tapered filtering only entails lower-order pointwise operations. Likewise, each Stage-2 iteration involves $N_\mathrm W$ \acp{IFFT}, one \ac{FFT}, and $\mathcal O(K_\mathrm p)$ pointwise operations. Since $N_\mathrm W=3$ is fixed, the overall complexity scales as
\begin{equation}
\mathcal O\!\left(
(T_{\max}+1)K_\mathrm p\log_2 K_\mathrm p
\right)
\label{eq:prop_complexity}
\end{equation}
where $T_{\max}$ is selected offline according to the spectral gap configuration. Hence, for any fixed gap configuration, the proposed method is independent of the number of targets, whereas the \ac{OMP} complexity grows with $L$. This makes the proposed approach a scalable alternative whose complexity is lower than \ac{OMP} and independent of the target count for a fixed gap configuration.
%where $T_{\max}$ denotes the number of Stage-2 iterations. As already mentioned, $T_{\max}$ is selected offline according to the spectral gap configuration and generally increases with the gap size. However, for any fixed gap configuration, $T_{\max}$ remains constant as the number of targets increases. Since the \ac{OMP} complexity instead scales with $L$, the proposed method provides a target-count-independent and scalable low-complexity alternative.
%
\begin{figure*}[t]
    \centering
    \includegraphics[width=0.76\linewidth]{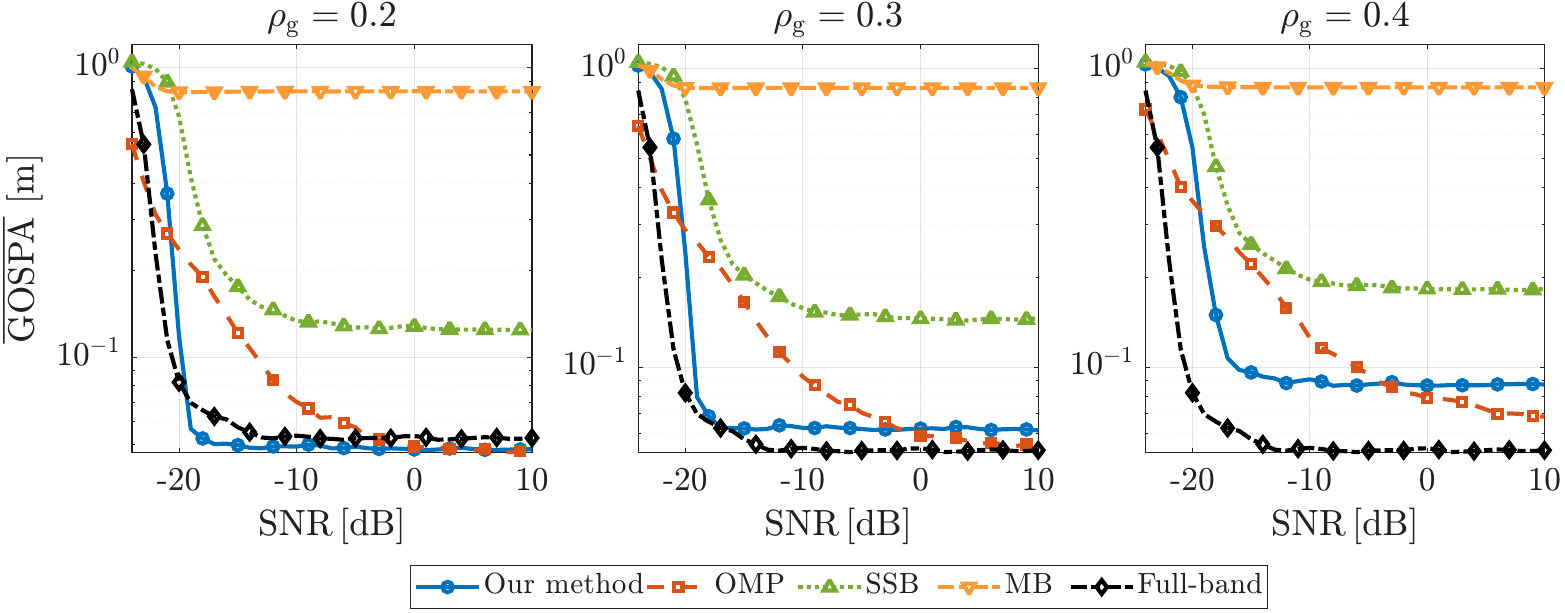}
    \caption{Mean GOSPA as a function of the $\mathrm{SNR}$ when $N_\mathrm{s} = 6$ targets are present.}
    \label{fig:GOSPA_vs_SNR}
\end{figure*}
%===========================================================
\section{Numerical Results}
\label{sec:num_res}
%===========================================================
This section evaluates the robustness of the proposed iterative reconstruction method in challenging range-only scenarios with multiple equal-power and closely spaced reflections. %System parameters representative of the upper mid-band \ac{FR}3 are considered for the \ac{OFDM} signal \cite{Bazzi:J26}. In particular, the carrier frequency is $f_\mathrm{c}=10$~GHz, the subcarrier spacing is $\Delta f=30$~kHz, the OFDM symbol duration is $T_\mathrm{s}=35.67\,\mu$s, and the total number of active subcarriers is $K=16380$, corresponding to an overall bandwidth of $B_\mathrm{max}\approx 500$~MHz. 
We consider upper mid-band FR3 \ac{OFDM} parameters \cite{Bazzi:J26}, namely $f_\mathrm{c}=10$~GHz, $\Delta f=30$~kHz, $T_\mathrm{s}=35.67\,\mu$s, and $K=16380$ active subcarriers, corresponding to $B_\mathrm{max}\approx 500$~MHz.
To isolate range-domain effects, transmit and receive beamforming are assumed already aligned toward a given sensing direction, where $N_\mathrm{s}$ point targets are present. Without loss of generality, the targets are assumed to have equal normalized power, i.e., $|\gamma_l|^2=1$ for all $l$, and are uniformly distributed over the range interval $R\in[20,85]$~m, with a minimum mutual separation of $2\Delta r_\mathrm{full}$, so as to avoid overlaps that would be irresolvable even in the full-band case. Under these assumptions, the same post-beamforming target \ac{SNR} is enforced for all scatterers according to \eqref{eq:SNR} by properly setting the noise variance $\sigma_\nu^2$.

We assess the range-estimation performance as a function of: i) the number of scatterers $N_\mathrm{s}$ for $\mathrm{SNR}=-5$~dB (see Fig.~\ref{fig:GOSPA_vs_Ntg}); and ii) the \ac{SNR} for a fixed $N_\mathrm{s}=6$ (see Fig.~\ref{fig:GOSPA_vs_SNR}). %\footnote{The \ac{SNR} is intended per target and is assumed to be the same for all targets.} 
Both analyses are carried out for three values of the frequency-gap fraction, namely $\rho_\mathrm{g}\in\{0.2,0.3,0.4\}$. The proposed method is compared with: i) the \ac{OMP} benchmark; ii) \ac{SSB} sensing, where only one sensing subband is used; iii) multiband sensing with null-gap filling (denoted as \ac{MB}), i.e., without reconstruction; and iv) the full-band case as reference.

The following algorithmic parameters are used. For Stage~1, Algorithm~\ref{alg:adaptive_equalizer} employs $\eta_\mathrm{dB}=-20$~dB, $J_\mathrm{max}=10$, and $c_w=2\sqrt{2\ln 2}\approx 2.3548$, with peak detection implemented through MATLAB's \texttt{findpeaks}. For Stage~2, the maximum number of iterations is set to $T_\mathrm{max}=22$, $28$, and $39$ for $\rho_\mathrm{g}=0.2$, $0.3$, and $0.4$, respectively. In both stages, the reconstruction grid size is set to $K_\mathrm{p,reb}=2^{18}$. For the final range-profile evaluation and for the \ac{OMP} correlation step, we use $K_\mathrm{p,est}=K_\mathrm{p,omp}=2^{15}$. The same zero-padding factor is also adopted for \ac{SSB}, \ac{MB}, and full-band sensing, so as to ensure the same range-grid spacing $\Delta r_\mathrm{grid}=0.15$~m for all compared methods.

To jointly account for localization error, missed detections, and ghost targets generated by the pronounced grating lobes induced by the spectral gap, we use the \ac{GOSPA} metric, which is widely adopted in multi-target localization. Specifically, we employ the standard \ac{GOSPA} formulation in \cite[Eq.~(12)]{Mat25}, with gating parameter $\xi_g=5\Delta r_\mathrm{full}\approx 1.5$~m. The reported results correspond to the mean metric $\overline{\mathrm{GOSPA}}$, obtained by averaging the \ac{GOSPA} values computed for each realization over $1000$ Monte Carlo runs. Range detection and estimation are performed on range power profiles normalized to unit peak in linear scale, through peak picking using MATLAB's \texttt{findpeaks} with \texttt{MinPeakProminence}$=0.1$, except for \ac{OMP}, for which the detected delays are directly returned. The results in Fig.~\ref{fig:GOSPA_vs_Ntg} and Fig.~\ref{fig:GOSPA_vs_SNR} show that \ac{GOSPA} generally increases with $N_\mathrm{s}$ because of stronger inter-target interference, even when the reconstruction itself remains accurate. In this respect, the proposed method exhibits a clear robustness advantage: for $\rho_\mathrm{g}\leq 0.3$, it remains close to the full-band reference as $N_\mathrm{s}$ increases, and for $\rho_\mathrm{g}=0.2$ it even slightly improves upon it thanks to its sidelobe-reduction effect. A more marked degradation appears only for $\rho_\mathrm{g}=0.4$, where the larger spectral gap makes the reconstruction less accurate on average and increases the probability of residual grating lobes and ghost detections. By contrast, \ac{OMP} is more sensitive to both the number of scatterers and the gap size, with the largest degradation observed in the practically relevant low-\ac{SNR} regime. Lastly, \ac{SSB} and null-gap \ac{MB} sensing provide the worst performance because of bandwidth loss in the former case and unresolved grating lobes in the latter. Overall, the results show that the proposed method effectively mitigates the impact of the spectral gap while preserving good multi-target range-estimation performance with a target-count-independent complexity for a fixed gap configuration.

\section{Conclusion}
\label{sec:conclusion}
This paper addressed coherent multiband \ac{OFDM} sensing in an intra-band configuration with two sensing subbands separated by a spectral gap. To cope with the resulting missing-subcarrier problem and the associated grating lobes, a low-complexity iterative reconstruction method was proposed, combining delay-domain equalization and apodization-based refinement, enforcing data consistency. The results showed that the proposed approach can achieve, for moderate gap sizes, multi-target range-estimation performance close to the full-band reference, while avoiding the target-count-dependent complexity growth of \ac{OMP}. These findings indicate that structured gap reconstruction makes coherent multiband sensing a practical and effective solution in scenarios where a wide contiguous sensing band cannot be allocated.

\bibliographystyle{IEEEtran}
\bibliography{IEEEabrv,bibliography}

\end{document}